\documentstyle[aps,epsf]{revtex}
\begin{document}
\title { Proton Spin Based On Chiral Dynamics}
\author{H. J. Weber }  
\vspace{0.3cm}
\address{Institute of Nuclear and Particle Physics,
University of Virginia, \\Charlottesville, VA 22903, USA }
\maketitle

\vspace{0.5cm}
\centerline{\bf Abstract}

\begin{abstract}
Chiral spin fraction models agree with the proton spin data only when  
the chiral quark-Goldstone boson couplings are pure spinflip. 
For axialvector coupling from soft-pion physics this is true for massless 
quarks, and at high momentum for light quarks. Axialvector quark-Goldstone 
boson couplings with {\em constituent} quarks are found to be 
{\em inconsistent} with the proton spin data.  
\end{abstract}
\vspace{0.5cm}
\centerline{Talk presented at the Symposium In Memory Of Judah M. Eisenberg}

\section{Introduction}

The nonrelativistic quark model (NQM) explains qualitatively many of the 
strong, electromagnetic and weak interaction properties of the nucleon and 
other octet and decuplet baryons in terms of three valence quarks whose 
dynamics is motivated by quantum chromodynamics (QCD), the gauge field theory 
of the strong interaction. Due to the spontaneous chiral symmetry breakdown 
($\chi $SB) of QCD, the effective degrees of freedom at the scale 
$\Lambda _{QCD}$ are expected to be quarks and Goldstone bosons.   

Here we point out that a naive use of a constituent quark mass for all 
observables, and in axialvector quark-Goldstone boson couplings in particular, 
leads to disagreement with the proton spin 
fractions, because the non-spinflip contributions dominate over spinflip 
at low momentum. 

Understanding the internal proton structure is one of the goals of particle 
physics and, over the past dozen years or more, has led to extensive studies 
of the spin and flavor contents of the nucleon in terms of measured deep 
inelastic structure functions (DIS)~\cite{SMC,E143}. Since spin fractions 
ultimately must derive from 
and be consistent with polarized DIS proton structure functions that are 
integrated over Bjorken $x$, we have used such a general DIS 
formalism~\cite{HJW} to analyze recent chiral models that have succeded in 
reproducing the spin fractions of the proton.      
\par    
The effects of chiral dynamics on 
the spin fractions are discussed in Sects. II and III in the framework of
chiral field theory applied to deep inelastic scattering.  


\section{Quark Spin Fractions from Chiral Dynamics}

Chiral field theory involves the effective strong interactions commonly used 
in chiral perturbation theory ($\chi$PT~\cite{CPT}) and applies at scales from 
$\Lambda _{QCD}$ up to the chiral 
symmetry restoration scale $\sim\Lambda _{\chi }=4\pi f_{\pi }\sim 1.17$ GeV,  
where $f_{\pi }=0.093$ GeV is the pion decay constant. 
\par
If the chiral symmetry breakdown is based on $SU(3)_L \times SU(3)_R$,
then the effective interaction between quarks and the octet of
Goldstone boson (GB) fields $\Phi_i$ involves the axial vector coupling    
\begin{eqnarray}
{\cal L}_{int}=-{g_A\over 2f_\pi }\sum_{i=1}^{8}\bar q \partial_\mu 
               \gamma ^\mu \gamma _5 \lambda _i \Phi_i q             
\label{lint}
\end{eqnarray}
that is well known from soft-pion physics. In Eq.~\ref{lint}, 
$g_A$ is the dimensionless axial vector-quark coupling constant 
that is taken to be 1 here. As a consequence, the polarization of quarks flips 
in chiral fluctuations, $q_{\uparrow,\downarrow,}\rightarrow 
q_{\downarrow,\uparrow}+GB$, into pseudoscalar mesons of the SU(3)
flavor octet of Goldstone bosons, but for massive quarks the
non-spinflip transitions from $\gamma_{\pm}\gamma_5 k^{\pm}$ that
depend on the quark masses are not
negligible. Let us also emphasize that, despite the nonperturbative 
nature of the chiral symmetry breakdown, the interaction between quarks and 
Goldstone bosons is small enough for a perturbative expansion to apply. 

\par 
Chiral field theory dissolves a dynamical or constituent quark into a 
current quark and a cloud of virtual Goldstone bosons. In this
context, it was first 
shown in~\cite{EHQ} that chiral dynamics leads to a reduction of the 
proton spin fractions carried by the valence quarks and also to a
reduction of 
the axial vector coupling constant $g_A^{(3)}$, based on one overall
chiral strength parameter, $a$, that contains the scale (cf. Table 1). It is 
well known that relativistic effects reduce the axial charge further, and this 
causes problems for the spin fractions~\cite{WB}. In addition, the violation 
of the Gottfried sum rule~\cite{Go}, which signals an isospin asymmetric
quark sea in the proton, i.e. $\bar u<\bar d$, became plausible. SU(3) 
symmetric chiral spin fraction models explain spin and sea quark observables 
of the proton, except for the ratio of axial charges $\Delta_3/\Delta_8=$5/3 
and the weak axial vector coupling constant of the nucleon,
$g_A^{(3)}={\cal F}+{\cal D}$. In~\cite{SMW,WSK} the effects of SU(3)
breaking (adding $(1+\lambda _8\epsilon )$ in ${\cal L}_{int}$) were more 
systematically 
built into these chiral models and shown to lead to a remarkable further 
improvement of the spin and quark sea fractions in comparison with the data.  
\par
It was first shown in ref.~\cite{WSK}, and subsequently
confirmed~\cite{XS,OS} that the $\eta '$ meson, proposed in~\cite{CLi}
mainly to decrease the antiquark fraction $\bar u/\bar d$ from the
SU(3) symmetric value $3/4$ to $\sim 0.5$, gives an almost negligible 
contribution to the spin fractions of the
nucleon, not only because of its large mass but also due to the small 
singlet chiral coupling constant. It is therefore often ignored, and this is
consistent with the understanding that, due to the axial anomaly, 
the $\eta '$ meson is not a genuine Goldstone boson. Pions and kaons
are well established
Goldstone bosons. For a discussion of the controversial role of the $\eta$
meson as the hypercharge or octet Goldstone boson, see~\cite{KW}.  
\par

Chiral fluctuations occur with probability densities 
$f(u_{\uparrow} \rightarrow \pi^+ + d_{\downarrow})$,... 
which, from Eq.~\ref{lint}, may be written as coefficients in the following 
chiral reactions: 
\begin{eqnarray}\nonumber 
u_{\uparrow} &\rightarrow& f_{u\rightarrow \pi ^{+}d}(x_{\pi }, 
\vec{k}_{\perp}) (\pi^+ + d_{\downarrow})
+f_{u\rightarrow \eta u} {1\over 6}(\eta +u_{\downarrow})
+f_{u\rightarrow \pi ^0 u} {1\over 2}(\pi^0 + u_{\downarrow})
\\\nonumber
&&+f_{u\rightarrow K^{+}s} (K^+ + s_{\downarrow}),\\\nonumber
d_{\uparrow} &\rightarrow&  f_{d\rightarrow \pi ^{-} u} (\pi^- + u_{\downarrow})
+f_{d\rightarrow \eta d} {1\over 6}(\eta +d_{\downarrow})
+f_{d\rightarrow \pi ^0 d} {1\over 2}(\pi^0 + d_{\downarrow})\\\nonumber
&&+f_{d\rightarrow K^0 s} (K^0 + s_{\downarrow}),\\
s_{\uparrow} &\rightarrow &f_{s\rightarrow \eta s} {2\over 3}(\eta 
   +s_{\downarrow})+f_{s\rightarrow K^- u} (K^- + u_{\downarrow})
+f_{s\rightarrow \bar K^0 d} (\bar K^0 + d_{\downarrow}),  
\label{fluc}
\end{eqnarray}
and corresponding ones for the other initial quark helicity. The
factors $1/2,1/6,1,...$ in Eq.~\ref{fluc} for $u\rightarrow \pi ^0 u, 
u\rightarrow \eta u, u\rightarrow \pi ^{+}d,...$, respectively,
originate from the flavor dependence in Eq.~\ref{lint} and are denoted as 
$p_m$ for the Goldstone boson $m$ for brevity. After integrating 
over transverse momentum in the infinite momentum frame, the coefficients 
in Eq.~\ref{fluc} become the polarized ($-$ sign) and unpolarized ($+$ sign) 
chiral splitting functions, 
\begin{eqnarray}
P^{\pm}_{GB/q}(x)=\int d^2 k_{\perp}f^{\pm}_{q\rightarrow q'GB}
(x,\vec{k}_{\perp}).
\label{splf} 
\end{eqnarray}
The unpolarized splitting function $P^+$ determines the (spinflip plus 
non-spinflip) probability for finding a Goldstone boson of mass
$m_{GB}$ carrying the 
longitudinal momentum fraction $x_{GB}$ of the parent quark's momentum and a 
recoil quark $q'$ with momentum fraction $1-x_{GB}$ for each fluctuation in 
Eq.~\ref{fluc}. 

\par
In (non-renormalizable) chiral field theory with cutoff $\Lambda _{\chi} $ of 
ref.~\cite{EHQ}, the unpolarized chiral splitting function takes the form 
\begin{eqnarray}
P^{+}_{q\rightarrow q'+GB}(x_{GB})={g^{2}_{A}\over f^{2}_{\pi }}
{x_{GB}\over 32\pi ^2}(m_q+m_{q'})^2\int^{t_{min}}_{-\Lambda _{\chi }^2}dt 
{(m_q-m_{q'})^2-t\over (t-m^{2}_{GB})^2}\ ,
\label{EHQ}
\end{eqnarray}    
where $t=k^2=-[(k_{\perp})^2+x_{GB}[(m'_{q})^{2}-(1-x_{GB})(m_{q})^2]]/
(1-x_{GB})$ is the square of the Goldstone boson four-momentum. The polarized 
splitting function is obtained using that it contains the difference of 
non-flip and helicity-flip probabilities. Since quarks are on their mass shell 
in the light front dynamics used here, the axialvector quark-Goldstone boson 
interaction is equivalent to the simpler $\gamma _5$ coupling. Except for an 
overall factor, the relevant unpolarized chiral transition probability is 
proportional to 
\begin{eqnarray}
-{1\over 2}tr[(\gamma \cdot p+m_q)\gamma _5(\gamma \cdot p+m_q)\gamma _5]
=2(pp'-m_q m'_q)=(m_q-m'_q)^2-k^2,\
\label{unpol}
\end{eqnarray}
where $2pp'=m'^{2}_{q}+m^{2}_{q}-k^2.$ Eq.~\ref{unpol} is the numerator in 
Eq.~\ref{EHQ} which can also be written as   
\begin{eqnarray}
{1\over 1-x_{GB}}[(k_{\perp})^2+[m'_{q}-(1-x_{GB})m_q]^2], 
\label{nu}
\end{eqnarray}
and has the following physical interpretation. Recall that the axialvector 
quark-Goldstone boson coupling ${\gamma_{\mu }\gamma_5 k^{\mu}}$ in 
Eq.~\ref{lint} involves the spin raising and lowering operators
$\sigma_1\pm i\sigma_2$ in a scalar product with the transverse
momentum $\vec{k}_{\perp}$ of the recoil quark, which suggests that the 
$k^{2}_{\perp}$ term in $P^+$ of Eq.~\ref{nu} represents the 
helicity-flip probability of the chiral fluctuation, while the 
longitudinal and time components, ${\gamma_{\pm}\gamma_5 k^{\pm}}$, induce the 
non-spinflip probability, which depends on the quark masses. This can be seen 
from the helicity non-flip probability 
\begin{eqnarray}
|\bar u'_{\uparrow}\gamma _5 u_{\uparrow}|^2=|\bar u'_{\downarrow}\gamma _5
u_{\downarrow}|^2\sim (m'_q-x' m_q)^2,\ x'=1-x_{GB},
\label{nofli}
\end{eqnarray}
using light cone spinors and suppressing the spinor normalizations. Thus
Eq.~\ref{nofli} identifies the mass term in Eq.~\ref{nu} as the helicity 
non-flip chiral transition. Similarly, the helicity-flip probability is 
obtained from 
\begin{eqnarray}
|\bar u'{\downarrow}\gamma _5 u_{\uparrow}|^2\sim (p'_{\perp})^2+x'^2 
(p_{\perp})^2-x'p'_{\perp}\cdot p_{\perp}
\label{fli}
\end{eqnarray}
which, in frames where $\vec{p}_{\perp}=0,$ reduces to 
$(\vec{k}_{\perp})^2,$
i.e. the net helicity flip probability generated by the chiral splitting 
process. In the nonrelativistic limit, where $|\vec{p}'_{\perp}|\ll 
m'_{q},$ $|\vec{p}_{\perp}|\ll m_q,$ clearly non-spinflip dominates over 
spinflip, while spinflip dominates at high momentum (it is not clear how  
ref.~\cite{EHQ} reached the opposite conclusion which, obviously, flies in 
the face of an extensive low-energy nuclear physics lore).  
\par 
   The polarized splitting function $P^-$ therefore has the same quark 
mass dependence as $P^+$, but involves the helicity flip probability with the 
opposite sign, i.e. has   
\begin{eqnarray}
{1\over 1-x_{GB}}[-(\vec{k}_{\perp})^2+\left(m'_{q}-(1-x_{GB})m_q\right)^2]
\label{polnu}
\end{eqnarray} 
in its numerator, which agrees with ref.\cite{SW}.    
Only for massless quarks there are no non-spinflip chiral transitions, so that 
$P^-=-P^+$ holds which is characteristic of pure spinflip chiral
transitions. 
    
\par
The splitting of quarks into a Goldstone boson and a recoil quark corresponds 
to a factorization of DIS structure functions that leads to a 
convolution of quark distributions with splitting functions.  
Thus, chiral fluctuations in lowest order of perturbation 
theory contribute convolution integrals  
\begin{eqnarray}\nonumber
\sum_{q,m} p_m P^{+}_{u_{\uparrow}\rightarrow q_{\downarrow} m}\otimes 
u^{0}_{\uparrow}+\sum_{q,m} p_m P^{+}_{d_{\uparrow}
\rightarrow q_{\downarrow} m} \otimes d^0_{\uparrow},\\   
\sum_{q,m} p_m P^{+}_{u_{\downarrow}\rightarrow +q_{\uparrow} m} 
\otimes u^0_{\downarrow}\
+\sum_{q,m} p_m P^{+}_{d_{\downarrow}\rightarrow 
q_{\uparrow} m} \otimes d^0_{\downarrow}
\label{rqp}
\end{eqnarray}
to the quark distributions $q_{\downarrow,\uparrow}$, respectively. Chiral 
fluctuations also cause a reduction of the valence quark probabilities  
\begin{eqnarray}
(1-P_q)q^{0}_{\uparrow,\downarrow},  
\label{valq} 
\end{eqnarray}
where $P_q$ are the total fluctuation probabilities. 

\section{Spin Distribution Results} 

In chiral dynamics, antiquarks originate only from the Goldstone bosons via 
their standard quark-antiquark composition.  
Therefore, {\em antiquarks are unpolarized}. Small antiquark 
polarizations are consistent with the most recent SMC data~\cite{SMC}, 
so that we expect only 
small corrections if we use $\bar u_\uparrow = \bar u_\downarrow$ in the spin 
fractions $\Delta u = u_\uparrow -u_\downarrow +\bar u_\uparrow 
-\bar u_\downarrow$, etc., i.e. $\Delta s=\Delta s_{sea}$, 
$\Delta \bar u=\Delta \bar d=\Delta \bar s =0$. 

\par
Let us now return to the polarized quark distributions and their lowest 
moments, the spin fractions.   
\par
Upon generalizing the chiral spin fraction formalism of~\cite{WSK,XS,CLi} to 
the polarized quark distributions, the probabilities displayed in 
Eq.~\ref{fluc},\ref{rqp},\ref{valq} yield  
\begin{eqnarray}
q_{\uparrow}(x)=(1-P_{q})q^{0}_{\uparrow}(x)+\sum_{m,q'} p_m 
P^{+}_{q'\rightarrow q m}\otimes q'^{0}_{\downarrow}+...
\label{oldq} 
\end{eqnarray}
which obviously are based on pure spinflip chiral transitions.
The corresponding result holds for the other quark helicity. 
The ellipses in Eq.~\ref{oldq} denote double convolution terms with 
$q^{0}_{\uparrow}$ from a
Goldstone boson $m$ that cancel in $q_{\uparrow}-q_{\downarrow}$. The opposite 
quark helicity on the rhs of Eq.~\ref{oldq} implies the {\em negative} sign of 
all chiral contributions to the spin distributions
\begin{eqnarray}
\Delta q(x) = (1-P_{q})\Delta q^0(x) -\sum_{m,q'} p_m P^{+}_{q'\rightarrow q m}
\otimes \Delta q'^0. 
\label{delp}
\end{eqnarray}
This result is common to all recent successful chiral models of spin
fractions. Let us emphasize that the general reduction of the valence
quark spin fractions $\Delta q^{0}$ by chiral fluctuations in lowest
order in Eq.~\ref{delp} is the 
crucial property responsible for the remarkable success of chiral field theory 
for the proton spin fractions. Eq.~\ref{delp} can be compared to the 
corresponding one from DIS involving the polarized splitting functions  
\begin{equation}
P^{-}_{q\rightarrow q' GB}(y)=\int d^2 k_{\perp} f^{-}_{q\rightarrow q' 
GB}(y, \vec{k}_{\perp}), 
\label{polspl}
\end{equation}    
which has the same form as Eq.~\ref{delp},    
\begin{eqnarray}
\Delta q(x) = (1-P_q)\Delta q^0(x)+\sum_{m,q'}p_m P^{-}_{q'\rightarrow q m}
\otimes \Delta q'^0
\label{delq}
\end{eqnarray}
except for the replacement of $-P^+$ by the corresponding polarized splitting 
function $P^-$. A comparison with Eqs.~\ref{nu},\ref{polnu} shows that this 
approximation, $P^-\approx -P^+$, corresponds to neglecting the
non-spinflip probability, which is valid only if the quark mass term in the 
numerator (i.e. Eq.~\ref{polnu} of the splitting functions is negligible 
compared to the tranverse momentum scale. This is not the case for constituent 
quarks~\cite{HJW,SW}.   

Thus, when these $\Delta q(x)$ of Eq.~\ref{delq} are integrated over Bjorken 
$x$, the lowest moments reproduce precisely the structure of the results for 
the spin fractions~\cite{WB,WSK} (cf. Table 1).  
\begin{table}[tbh]
\caption{\label{tab:1} Quark Spin Observables of the Proton (from 
ref.~\protect\cite{WSK}), $a=$chiral strength, $\zeta =$relative 
singlet to octet coupling and $\epsilon =SU_3$ breaking parameter in 
$\lambda _8$ direction.}
\begin{tabular}{|c|c|c|c|c|}
\hline
& Data            & NQM & $a=$0.12 & $a=$0.12\\ 
& E143\cite{E143} &     & no $\eta '$  & with $\eta '$  \\
& at 3 GeV$^2$    &     &       & $\zeta=$-0.3 \\
& SMC\cite{SMC} &       & $\epsilon =$ 0.2 & $\epsilon =$ 0.2 \\
& at 5 GeV$^2$    &     &          &          \\
\hline
$\Delta u$ & 0.84$\pm$ 0.05 & 4/3 & 0.83 & 0.81  \\
           & 0.82$\pm$0.02  &     &      &       \\
$\Delta d$ & -0.43$\pm$0.05 & -1/3& -0.39& -0.39 \\
           & -0.43$\pm$0.02 &     &      &     \\
$\Delta s$ & -0.08$\pm$0.05 & 0 & -0.07 & -0.07\\
           & -0.10$\pm$0.02  &   &       &   \\    
$\Delta \Sigma $ & 0.30$\pm$0.06 & 1 & 0.36 & 0.35 \\
            & 0.29$\pm$0.06 & & & \\       
$\Delta_3/\Delta_8$ & 2.09$\pm$0.13 & 5/3 & 2.12 & 2.13 \\
$g^{(3)}_A$  & 1.2573$\pm$0.0028\cite{PDG}& 5/3 & 1.22 & 1.21 \\
${\cal F}/{\cal D}$ & 0.575$\pm$0.016 & 2/3 & 0.58 & 0.58 \\
$I_G$      & 0.235$\pm$0.026 & 1/3 & 0.27 & 0.25 \\
\hline
\end{tabular}
\end{table}
\section{Conclusions}  
\par
The detailed comparison in Sect. IV shows unambiguously that the success of 
several recent chiral models~\cite{EHQ,WB,SMW,WSK,XS,OS,CLi} for the spin 
fractions $\Delta q$  can be attributed to pure spinflip quark-Goldstone 
boson couplings that do not account for quark mass terms from the non-spinflip 
chiral transitions in the lowest moments of the splitting functions in 
standard chiral field theory. 

Whenever {\em constituent quark masses} are used in
Eqs.~\ref{nu},\ref{polnu}, then the positive contributions in $P^-$ 
corresponding to the non-spinflip probability represented by the quark mass 
term substantially reduce the chiral
subtractions so that {\em no agreement with the proton spin data can
be achieved}~\cite{HJW,SW}. These results were obtained with quite 
different initial valence quark distributions and both show that the
proton spin observable $\Delta \Sigma $ stays above the value 1/2. This 
disagreement can be interpreted so that the constituent quark model in the 
framework of standard chiral field theory, which is often called chiral quark 
model~\cite{GM}, is ruled out by the proton spin data.     
\par
The success of the chiral spin fraction models suggests that axialvector 
quark-Goldstone boson couplings are consistent with current quarks, but not 
constituent quarks. This usage conforms with the chiral field theory practiced 
in chiral perturbation theory~\cite{CPT}.  
It is interesting to note that proton spin data are 
successfully described by models based on an instanton fluid in the vacuum, 
where the instanton-quark interaction is also pure spinflip~\cite{DK}. 

The proton spin data does not challenge the vast nuclear theory based on pion 
exchange directly -- as the pion exchange potential has been successfully 
tested through its pion exchange current predictions -- but it suggests 
understanding better its derivation from QCD concepts such as current quark 
masses and quark condensates, and not merely from constituent quark models.

\end{document}